# REMOTE DEVICE ACCESS IN THE NEW CERN ACCELERATOR CONTROLS MIDDLEWARE

N. Trofimov, IHEP, 142284, Protvino, Russia
V. Baggiolini, S. Jensen, K. Kostro, F. Di Maio, A. Risso, CERN, Geneva, Switzerland

## Abstract

This paper presents the Remote Device Access (RDA) package developed at CERN in the framework of the joint PS/SL Controls Middleware project. The package design reflects the Accelerator Device Model in which devices, named entities in the control system, can be controlled via properties. RDA implements this model in a distributed environment with devices residing in servers that can run anywhere in the controls network. It provides a location-independent and reliable access to the devices from control programs. By invoking the device access methods, clients can read, write and subscribe to device property values. We describe the architecture and design of RDA, its API, and CORBA-based implementations in Java and C++. First applications of RDA in the CERN accelerator control systems are described as well.

## 1 INTRODUCTION

In 1999 an initiative was launched to create a common software communication infrastructure for the CERN accelerator controls. This infrastructure should replace existing heterogeneous software protocols and components and provide new facilities for the LHC era, in particular:

- Support the standard Accelerator Device Model and device I/O services [1].
- Support the publish/subscribe paradigm and synchronization of application programs with Accelerator Timing.
- Provide inter-operability solutions for industrial control systems.

It has also been requested to use *available standards* and *commercial products*. Based on these requests, the Controls Middleware (CMW) project [2] was launched. Following the technology study and the requirements capture, the middleware technology was selected and the base architecture proposed. A number of software components implementing the proposed architecture were developed; one of them is the RDA package, which provides access to the accelerator devices from application programs in a distributed heterogeneous environment.

## 2 DESIGN CHOICES

Besides the already mentioned general requirements, the design of RDA was significantly influenced by the need to provide multi-language and multi-platform inter-operability. RDA should act as a "software bus" that transparently interconnects applications and devices implemented in different languages (Java, C++, C) and running on any of the platforms used in CERN accelerator controls (Linux, HP-UX, LynxOS, Windows).

The choice of CORBA as a communication technology looks obvious in this situation. It was decided, however, to restrict its use to the RDA internal software. All CORBA interface and data definitions are hidden in the package and do not appear in the RDA application programming interface (API).

It has been discussed for a long time whether a *wide* or *narrow* API would be more appropriate. Wide API (different calls for different classes) would expose CORBA to our users and allow them to define device classes and device-specific data types in CORBA IDL. The main advantage of this approach is strong compile-time type checking. Narrow API (same call for all classes) allows less type checking and imposes some restrictions on data types, but it can remain stable for a long time, and this solution was finally chosen in the RDA design. Our previous experience with RPC-based systems (Remote Procedure Call) played a role here: a big number of user-defined interfaces proved to be difficult to manage and was replaced by a single device access API.

To avoid compile-time dependency on specific device data types, RDA uses generic containers where values are passed between applications and devices, along with their run-time type descriptions. In this respect we have been influenced by CDEV [3], and the CDEV concept of the Data object has been adopted almost without change. The CORBA *any* type is used to transport contents of the Data objects.

## 3 ARCHITECTURE

RDA is based on a client-server model. Accelerator devices are implemented in device servers, and client applications access them using the classes and interfaces provided in the RDA client API (Figure 1).

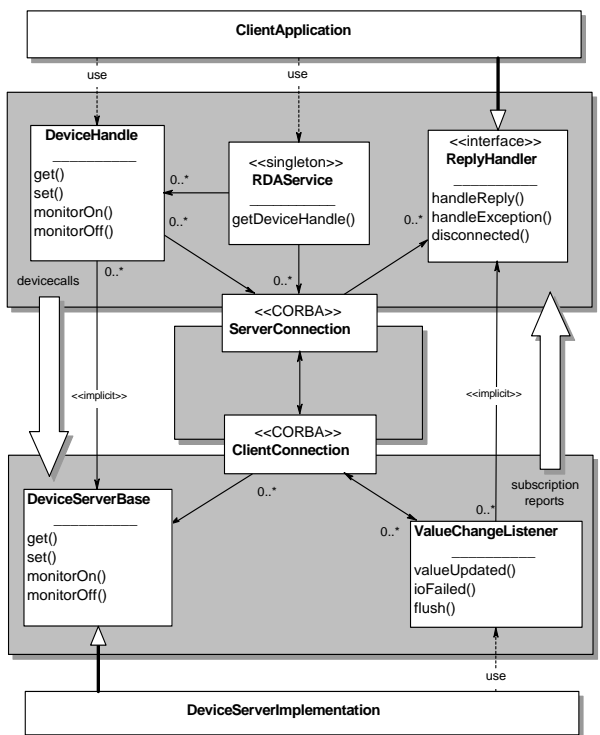

Figure 1: Architectural overview of RDA.

The *RDAService* class manages communications with remote devices and serves as a factory for *DeviceHandle* objects. A client application uses a *DeviceHandle* to remotely invoke access methods on a device. The *DeviceHandle* delegates device calls over the transport layer to the *DeviceServerBase* object that represents the device server in which the "target" device resides. *DeviceServerBase* is an abstract class that provides the RDA connectivity for concrete device servers, which are implemented as its derived classes. A device server developer should at least implement the four abstract methods declared in the *DeviceServerBase* class: these methods are called whenever a corresponding method is invoked on an associated device handle.

The *get* and *set* methods can be invoked either in synchronous (blocking) or asynchronous (non-blocking) mode. All asynchronous calls require a reference to the object implementing the *ReplyHandler* interface as a parameter; when a reply to the request arrives at the client, the RDA will pass the operation results to the specified object using methods defined in this interface.

An object that implements the *ReplyHandler* interface must also be specified in each *monitorOn* call. This object will receive subscription reports from the associated *ValueChangeListener* objects on the server side. For each incoming subscription request, the server-side RDA creates a *ValueChangeListener* object. The implementation uses this object to forward updates of property value or errors to the client (via the clients ReplyHandler object).

The transport layer employs the concept of "connection" between a client and a server. The connection is represented by the ServerConnection class on the client side and by the ClientConnection class on the server side. These two classes provide all functions needed to send operation requests and receive replies, as well as functions to control the actual network connection which they represent. The "connection" classes are implemented as CORBA objects, thus allowing the object representing one end of the connection to invoke remotely methods of the peer object on the other end.

The use of the transport level connection in RDA is asymmetric: a server provides an "entry point" for the connection establishment on the initiative of a client wanting to send a request to that server. In the CORBA implementation, clients use the Naming Service to obtain references to CORBA objects that represent such entry points in the device servers. A connection is normally closed on the initiative of a client. Abnormal disconnections (e.g., due to a client or a server crash) are detected by the RDA connection monitoring mechanism which is based on "pinging" from a client to a server. When a client dies, the server no longer receives the ping, and releases all local resources related to this client.

If a ping or an ordinary operation fails in a connection due to an irrecoverable communications error, the client-side *ServerConnection* concludes that the server is inaccessible and invokes the *disconnected* method on all reply handlers waiting for subscription reports. It then starts to monitor the server reference on the Naming Service. When the server is up again, it reregisters with the Naming Service; this triggers the reconnection procedure: the client will attempt to reconnect to the server and resend all pending subscription requests. The connection monitoring and recovery are performed internally by the RDA and are fully transparent to the client and server applications.

## 4 CORBA PRODUCT SELECTION

The RDA uses only standard CORBA 2.2 facilities, so that any standard compliant ORB can be used in the RDA transport layer. Although CORBA products are now available from a large number of vendors, selection of a product that meets all our requirements turned out not to be not easy. It was especially difficult to find a suitable, fast and "lightweight" ORB for LynxOS front-ends, where resources are limited. We evaluated a number of CORBA products, both commercial and public domain, and finally selected the ORBacus family of products from IONA, in particular ORBacus/E, which is targeted at embedded real-time applications. ORBacus/E is a commercial product but it is free for non-commercial use and available in source code. LynxOS is not an officially supported platform for ORBacus, but the port to LynxOS 3.0 and 3.1 was relatively straightforward and did not require any significant modifications in the source code.

## 5 CURRENT STATUS

A full implementation of RDA in Java has been available for about a year now. The C++ server part is already available on LynxOS, Linux and Windows platforms, while the client part implementation is in progress. In this chapter we will briefly describe two applications of RDA in CERN accelerator controls. Both applications use the Device Server Framework, which is another CMW product that extends the server-side RDA with a set of utility classes facilitating device server development.

### 5.1 RDA in AD Controls

A part of the PS complex, the Antiproton Decelerator (AD), is operated by means of Java application programs which use CDEV as the equipment access interface [1]. Until now, gateways were used to communicate with the equipment; these gateways are now being replaced by direct connections to RDA servers deployed on each front-end computer (Figure 2).

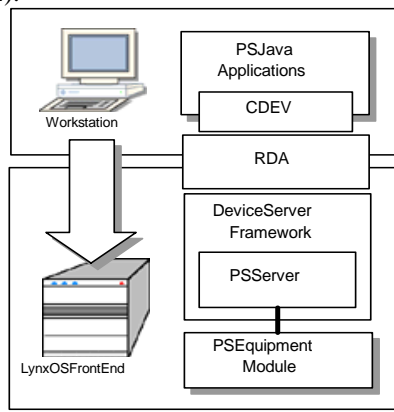

Figure 2: RDA in AD controls.

As a result, the equipment access performance has been significantly improved. A synchronous call from an application running on an 800 MHz PC to a device server on a 166 MHz PowerPC via 10 Mb Ethernet takes about 5 ms. Each device server can send to clients up to 1000 subscription updates every 1.2s (the accelerators' basic period). This limit is imposed mostly by the data acquisition speed in the server: the time required for RDA to transmit 1000 updates from a server to a client is less then 50 ms.

By interfacing CDEV with the RDA client, the actual AD programs can remain unchanged. This approach has been successfully tested and it will be used at the next AD startup.

### 5.2 OPC Gateway

Connection to industrial systems has been an important requirement of the CMW project. The selected solution was to use OPC, which is the de-facto standard in industrial controls. A server was developed which is using RDA and the CMW Server Framework. This server can connect to any *OPC Server* and maps each device/property to an OPC Item. This is done using an ORACLE database description so that each server can auto-configure. OPC itself supports subscriptions, which maps easily to RDA *monitor/update*. Three different SCADA system products have been connected this way.

## CONCLUSIONS

The RDA package implements data subscription and automatic reconnection facilities on top of standard CORBA, while hiding the CORBA complexity from users. CORBA allowed us to avoid much development, by using commercial products available from many vendors.

RDA implements synchronous get/set as well as publish/subscribe models and performances obtained in both are satisfactory.

We are currently deploying RDA-based servers on various platforms and accessing accelerator devices of different origins (PS, SL and industrial) with Java programs.

## REFERENCES

[1] P. Charrue, J. Cuperus, I. Deloose, F. Di Maio, K. Kostro, M. Vanden Eynden, "The CERN PS/SL Controls Java Application Programming Interface", ICALEPCS'99, Trieste, Italy, Oct. 4'8, 1999.

[2] http://proj-cmw.web.cern.ch/proj-cmw: The Web page of the Controls MiddleWare Project.

[3] J. Chen, G. Heyes, W. Akers, D. Wu, W. Watson, "CDEV: An Object-Oriented Class Library for Developing Device Control Applications", ICALEPCS'95, Chicago USA, Oct. 29 – Nov. 3, 1995.